\documentclass[
 reprint,
 superscriptaddress,
 amsmath,amssymb,
 aps,
floatfix,
pra]{revtex4-1}

\usepackage{graphicx}% Include figure files
\usepackage{dcolumn}% Align table columns on decimal point
\usepackage{bm}
\usepackage{mje}
\usepackage{hyperref}
\usepackage[utf8]{inputenc}
\usepackage{upgreek}
\usepackage{tikz}
\usetikzlibrary{positioning,shapes}

\newcommand{\su}[1] {\ensuremath{\mathfrak{su}\!\left( #1 \right)}\xspace}

\def\Bid{{\mathchoice {\rm {1\mskip-4.5mu l}} {\rm
{1\mskip-4.5mu l}} {\rm {1\mskip-3.8mu l}} {\rm {1\mskip-4.3mu l}}}}

\def\Balph{\boldsymbol{\alpha}}
\def\BalphT{\tilde{\boldsymbol{\alpha}}}
\def\CKHW{\hat{\mathsf{\Pi}}} % composite kernel HW
\def\CDHW{\mathsf{\OpD}} % composite displacement HW
\def\SpinPi{\hat{\Pi}}

\def\WigParamSU2{\Omega}
\def\WeyParamSU2{\Weyling{\Omega}}

\def\WigParamCompSU2{\mho}
\def\WeyParamCompSU2{\Weyling{\mho}}

\def\DO{\hat{\rho}}

\newcommand{\phiT}{\Weyling{\phi}}
\newcommand{\thetaT}{\Weyling{\theta}}
\newcommand{\PhiT}{\Weyling{\Phi}}
\newcommand{\CPSpace}{{\mathcal{C}\text{P}^{N-1}}}

\newcommand{\Parity}{\ensuremath{\hat \Pi}\xspace}

\newcommand{\Weyling}[1]{\tilde{#1}} % how the Weyl Coordinates come out
\newcommand{\GenKer}{\hat{\Delta}}
\newcommand{\arbitraryKernel}{\hat{\Delta}_s}
\newcommand{\arbitraryKernelN}[1]{\hat{\Delta}_{s_{#1}}}
\newcommand{\inverseKernel}{\hat{\Delta}^\dag_{-s}}
\newcommand{\inverseKernelN}[1]{\hat{\Delta}^\dagger_{-s_{#1}}}
\newcommand{\arbitraryParam}{\Theta}
\newcommand{\arbitraryCF}[1]{F_{\hat{#1}}}
\newcommand{\GenWigF}[1]{W_{\hat{#1}}}
\newcommand{\GenWeyF}[1]{\tilde{W}_{\hat{#1}}}
\newcommand{\GenWigKer}{\hat{\Pi}}
\newcommand{\GenWeyKer}{\hat{\mathcal{D}}}
\newcommand{\GenWigPar}{\Omega}
\newcommand{\SUNWigPar}{\Omega}
\newcommand{\CompWigPar}{\Omega}
\newcommand{\GenWeyPar}{\Weyling{\GenWigPar}}
\newcommand{\SUNWeyPar}{\Weyling{\SUNWigPar}}
\newcommand{\CompWeyPar}{\Weyling{\CompWigPar}}
\newcommand{\GenArg}{\OpA}

\newcommand{\GenFourier}{\mathcal{F}}
\newcommand{\GenConvolution}{\mathcal{K}}

\newcommand{\QSERG}{Quantum Systems Engineering Research Group, Department of Physics, Loughborough University, Leicestershire LE11 3TU, United Kingdom}
\newcommand{\Wolfson}{The Wolfson School, Loughborough University}

\begin{document}

\title{A general approach to quantum mechanics as a statistical theory}

\author{R.~P.~Rundle}
% Co-developed maths and physics for all sections as well as all simulations and figures. Noticing the Weyl function needed the full displacement operator. Figures. Applications
\affiliation{\QSERG}
\affiliation{\Wolfson}
\author{Todd Tilma}
% Co-developed/supported maths and physics for all sections as well as appendices.
\affiliation{Tokyo Institute of Technology, 2-12-1 \=Ookayama, Meguro-ku, Tokyo 152-8550, Japan}
\affiliation{\QSERG}
\author{J.~H.~Samson}
% Critical analysis, verification using toy models, and consideration of corner cases. Applications
\affiliation{\QSERG}
\author{V.~M.~Dwyer} 
% Ambiguity function connection. Help With Matlab Code, Applications
\affiliation{\QSERG}
\affiliation{\Wolfson}
\author{R.~F.~Bishop}
\affiliation{School of Physics and Astronomy, Schuster Building, The University of Manchester, Manchester, M13 9PL, United Kingdom}
\affiliation{\QSERG}
% Development and improvements in formalism and argument, Applications
\author{M.~J.~Everitt}
% Conceived original idea of Fourier transform of generalized displacements as a kernel for Weyl functions. Led project and writing of paper - formulation of main story/narative. Figures, Applications
\email{m.j.everitt@physics.org}
\affiliation{\QSERG}

\date{\today}

\begin{abstract}
Since the very early days of quantum theory there have been numerous attempts to interpret quantum mechanics as a statistical theory. 
This is equivalent to describing quantum states and ensembles together with their dynamics entirely in terms of phase-space distributions. 
Finite dimensional systems have historically been an issue. 
In recent works [Phys. Rev. Lett. 117, 180401 and Phys. Rev. A 96, 022117] we presented a framework for representing any quantum state as a complete continuous Wigner function. 
Here we extend this work to its partner function -- the Weyl function. 
In doing so we complete the phase-space formulation of quantum mechanics -- extending work by Wigner, Weyl, Moyal, and others to any quantum system. 
This work is structured in three parts. 
Firstly we provide a brief modernized discussion of the general framework of phase-space quantum mechanics. 
We extend previous work and show how this leads to a framework that can describe any system in phase space -- putting it for the first time on a truly equal footing to Schr\"odinger's and Heisenberg's formulation of quantum mechanics. 
Importantly, we do this in a way that respects the unifying principles of ``parity'' and ``displacement'' in a natural broadening of previously developed phase space concepts and methods. 
Secondly we consider how this framework is realized for different quantum systems; in particular we consider the proper construction of Weyl functions for some example finite dimensional systems. 
Finally we relate the Wigner and Weyl distributions to statistical properties of any quantum system or set of systems.
\end{abstract}

\maketitle

\section{Introduction}\label{Intro}

The field of quantum physics is undergoing rapid expansion, not only in such high-profile applications as those promised by quantum information technologies, but also in such foundational areas as quantum thermodynamics. 
Wigner was motivated by the latter context in his seminal work ``On the Quantum Correction For Thermodynamic Equilibrium''~\cite{Wigner1932}, where he defined the function that now takes his name. 
However, the original Wigner function, and its extensions~\cite{1601.07772, Rundle2016, WOOTTERS19871, Agarwal1981, Luis-052112, Luis-495302, Klimov055303, PhysRevA.86.062117, Kano1974}, are now finding great utility in the former context. 
The Wigner function is the quantum analog of the classical probability density which is a function of the system's state variables. In classical statistical mechanics there is another distribution which is of great importance, the characteristic/moment-generating function. These two classical distributions, being two-dimensional Fourier transforms of each other are, are naturally complementary and extremely powerful. There have been numerous attempts to bring to general quantum systems a similar framework - each of which have suffered from issues such as being informationally incomplete or being  singular in nature (see, for example,~\cite{Haken1967,Samson2000,Samson:2003ij,Scully1994,PhysRevA.6.2211}). In this work we describe how, by taking account of the underlying group structure, we can use a single general approach to quantum mechanics as a statistical theory that resolves these issues.

In many introductory texts, and even seminal works such as~\cite{Glauber1969, Wigner1984}, the Wigner function is introduced via the Weyl-Wigner transformation that describes transforming a Hilbert space operator \OpA to a classical phase-space function $\GenWigF{A}(q,p)$~\cite{Weyl1927, Ozizmir1967, Leaf1968-1, Leaf1968-2, Moyal1949}:
\bel{Weyl-Wigner-ND1}
\GenWigF{A}({q},{p}) = \int_{-{\infty}}^{+{\infty}} \ud {\zeta} \bra{{q}-\frac{{\zeta}}{2}} \OpA \ket{{q}+\frac{{\zeta}}{2}} \mathrm{e}^{\ui p\zeta/\hbar}. 
\ee 
Here $\iint \ud q \ud p \, \GenWigF{A}({q},{p}) \equiv 2\pi\hbar\Tr[\OpA]$ and we regain the function originally introduced by Wigner $\GenWigF{\rho}({q},{p})$ by replacing \OpA with the density operator $\DO$~\cite{Wigner1984}.
As a direct replacement of the density matrix, the Wigner function can serve to represent both pure and mixed states with the system dynamics described by a Liouville equation with quantum corrections~\cite{Moyal1949, Groenewold1946}. 
Thus it is possible to view the Wigner function as a quantum replacement of the probability density function in classical physics. 

In Wigner's original work~\cite{Wigner1932} the function of~\Eq{Weyl-Wigner-ND1} and its dynamics were introduced for a collection of particles,  
\bel{Weyl-Wigner-ND2} 
\GenWigF{\rho}(\mathbf{q},\mathbf{p})  =  \int_{-\boldsymbol{\infty}}^{+\boldsymbol{\infty}} \ud \boldsymbol{\zeta} \, \bra{\mathbf{q}-\frac{\boldsymbol{\zeta}}{2}} \DO \ket{\mathbf{q}+\frac{\boldsymbol{\zeta}}{2}} \, e^{i \mathbf{p} \cdot \boldsymbol{\zeta}/\hbar}, 
\ee 
where $\mathbf{q} = [q_1,q_2,\ldots,q_n]$ and $\mathbf{p} = [p_1,p_2,\ldots,p_n]$ are $n$-dimensional vectors representing the classical phase-space position and momentum values, and $\boldsymbol{\zeta} = [\zeta_1,\zeta_2,\ldots,\zeta_n]$ is a $n$-dimensional variable of integration.
Equation~(\ref{Weyl-Wigner-ND1}) results by integrating out the marginals of all but one component (in exactly the same way as one does a partial trace of a system's density operator)~\cite{Wigner1932}.

An equivalent method for generating a Wigner function of an ensemble can be done by performing a group action on the density matrix directly~\cite{Glauber1969, Groenewold1946},
\bel{WignerCompact}
\GenWigF{\rho}(\Balph) = 2^n \Trace{ \DO \, \hat{\mathcal{P}}(\Balph)}.
\ee
Here
\bel{alphaDef} 
\alpha_i = \frac{1}{\sqrt{2}}\left(\gamma_i q_i + \frac{\ui }{\gamma_i\hbar}p_i\right)
\ee
for $\gamma_i = \sqrt{m_i\omega_i/\hbar}$, and $\hat{\mathcal{P}}(\Balph) \equiv \bigotimes_i \hat{\mathcal{P}}_i(\alpha_i)$ is a displaced parity operator for the whole system.
This operator is built from the individual displaced parity operators, $\hat{\mathcal{P}}_i(\alpha_i) = \OpD_i(\alpha_i) \hat{\mathcal{P}}_i \OpD_i^\dag(\alpha_i)$, such that
\ba
\hat{\mathcal{P}}_i &\equiv & \exp \left(\ui \pi \Opa^\dag_i \Opa_i \right) \label{DefParityOp}\\
&=& \sum_{n=0}^{\infty} (-1)^{n}\ket{n}_{i\,i}\!\bra{n} \nonumber
\ea
is a diagonal operator basis of the eigenstates of the number operator ($\ket{n}_i$) and
\ba
\OpD(\alpha_i) &=& \exp\left(\alpha_i \Opad_i - \alpha_i^*\Opa_i \right) \label{displacementOp}\\
&\equiv& \exp\left(\ui[p_i\Opq_i - q_i\Opp_i \right]/\hbar) \nonumber
\ea
is the standard displacement operator~\cite{PhysRev.131.2766}. Here $\OpD(\alpha_i)$ is defined according to the annihilation and creation operators written in terms of the position, $\Opq_i$, and momentum, $\Opp_i$,  operators (with $\COM{\Opq_i}{\Opp_j}=\ui\hbar\,\delta_{ij}$) where
\bel{aDef}
\Opa_i = \frac{1}{\sqrt{2}}\left(\gamma_i\Opq_i + \frac{\ui }{\gamma_i\hbar}\Opp_i\right),
\Opad_i = \frac{1}{\sqrt{2}}\left(\gamma_i\Opq_i - \frac{\ui }{\gamma_i\hbar}\Opp_i\right),
\ee
so that $\COM{\Opa_i}{\Opad_j}=\delta_{ij}$.
Because we will later want to discuss general composite systems, we absorb the normalization of $2$ into the displaced parity operator to generate a normalized displaced parity operator
\be
\Parity_i(\alpha_i) \equiv 2 \hat{\mathcal{P}}_i(\alpha_i),
\ee
allowing us to rewrite~\Eq{WignerCompact} as
\ba 
\GenWigF{\rho}(\Balph) &=& \Trace{\DO \, \CKHW(\Balph)} \label{NormWignerHW}, \\
\CKHW(\Balph) &\equiv& \bigotimes_i \Parity_i(\alpha_i). \nonumber
\ea

When dealing with probability distribution functions, it is generally useful within a statistical framework to consider the corresponding characteristic function. 
The characteristic function has historically been given by the Fourier transform of the probability distribution function. 
In our case, taking the Fourier transform of the Wigner function yields the Weyl function~\footnote{We are following a similar argument outlined in Ref.~\cite{Samson2000} for the generation of the Weyl function.}
\be 
\GenWeyF{\rho}(\BalphT) = \left(\frac{1}{\pi}\right)^n \int^{+\boldsymbol{\infty}}_{-\boldsymbol{\infty}} \ud\Balph \, \GenWigF{\rho}(\Balph) \, \exp\left(\Balph\cdot\BalphT^* - \Balph^*\cdot\BalphT\right),
\ee
and similarly
\be 
\GenWigF{\rho}(\Balph) = \left(\frac{1}{\pi}\right)^n \int^{+\boldsymbol{\infty}}_{-\boldsymbol{\infty}} \ud\BalphT \, \GenWeyF{\rho}(\BalphT) \, \exp\left(\BalphT\cdot\Balph^* - \BalphT^*\cdot\Balph\right),
\ee
where $\Weyling{\alpha}_i$  is the dual of $\alpha_i$ such that $\Weyling{\alpha}_i = (\gamma_i\Weyling{q}_i+\ui\Weyling{p}_i/\gamma_i\hbar )/\sqrt{2}$. 
The Weyl function can be thought of as a $2n$-dimensional autocorrelation function, and so each $\Weyling{q}_i$ ($\Weyling{p}_i$) can be thought of as an increment of position (momentum). 
This is in the sense that they display the overlap between the state and the same state displaced by that position (momentum) increment.

This Weyl function~\cite{Wigner1984, Groenewold1946} was used by Moyal as a starting point in his work \emph{``Quantum Mechanics as a Statistical Theory''} and is a moment generating function of the quantum state or operator being considered~\cite{Moyal1949}.
The Weyl function can be defined in its own right in terms of a group action by
\bel{WeylCompact}
\GenWeyF{\rho}(\BalphT) = \Trace{\hat\rho \, \CDHW(\BalphT)},
\ee
where $\CDHW(\BalphT) \equiv \bigotimes_i \OpD_i(\tilde{\alpha}_i)$, and $\OpD_i(\tilde{\alpha}_i)$ is the displacement operator defined in~\Eq{displacementOp}.
To return the density matrix, the inverse transforms of~\Eq{NormWignerHW} and~\Eq{WeylCompact} are needed~\cite{Moyal1949, Glauber1969, Wigner1984}.  
This can be done by integrating the phase-space function with the Hermitian transpose of the kernel used to create that function~\cite{Glauber1969}.
\ba
\hat\rho &=& \left(\frac{1}{\pi}\right)^n \int^{+\boldsymbol{\infty}}_{-\boldsymbol{\infty}} \ud \Balph \; \GenWigF{\rho}(\Balph)\CKHW(\Balph) \label{HWWignerInverse}\\
\hat\rho &=& \left(\frac{1}{\pi}\right)^n \int^{{+\boldsymbol{\infty}}}_{{-\boldsymbol{\infty}}} \ud \BalphT \; \GenWeyF{\rho}(\BalphT) \CDHW^\dag(\BalphT). \label{HWWeylInverse}
\ea
Note that because parity is Hermitian the displaced parity must also be an Hermitian operator so that the adjoint is not needed in~\Eq{HWWignerInverse}.

%-----------------------------------------------------------------------------
%-----------------------------------------------------------------------------
\section{The General Framework}\label{SectionTwo}

\subsection{Phase-space distributions and their dynamics}
We have previously shown that it is possible to generalize the Wigner function to arbitrary systems~\cite{1601.07772}.
In this paper we will show that the same can be done for the Weyl function, yielding a complete and complementary representation of quantum mechanics in phase space. 
The general framework is described below with respect to any operator $\GenArg$.

To begin, consider an arbitrary phase-space function, ($\arbitraryCF{A}^{(s)}$) of $\GenArg$ defined with respect to a kernel which maps a state to phase space through a group action ($\arbitraryKernel$) parameterized over some phase space ($\arbitraryParam$). This can be written as 
\be
\arbitraryCF{A}^{(s)}(\arbitraryParam) = \Trace{\OpA\,\arbitraryKernel(\arbitraryParam)}.
\ee
Following Refs.~\cite{Glauber1969, PhysRevA.59.971}, the subscript $s$ in the kernel refers to the ordering of the operators: $1$ for normal, $0$ for symmetric, and $-1$ for anti-normal ordered (for those systems where this is meaningful; $s$ takes on alternative meaning for spins~\cite{PhysRevA.59.971}). 
When considering quasiprobability distribution functions, these values correspond to analogs of the Glauber-Sudarshan $P$ function ($s=1$)~\cite{PhysRev.131.2766, PhysRevLett.10.277}, the Wigner function ($s=0$)~\cite{Wigner1932}, and the Husimi $Q$ function ($s=-1$)~\cite{Husimi1940}.
%When considering quasiprobability distribution functions (i.e. where the kernel is a displaced parity operator), these values correspond to analogs of the Glauber-Sudarshan $P$ function ($s=1$)~\cite{PhysRev.131.2766, PhysRevLett.10.277}, the Wigner function ($s=0$)~\cite{Wigner1932}, and the Husimi $Q$ function ($s=-1$)~\cite{Husimi1940}.

Supposing that a suitable kernel exists~\cite{Glauber1969}, we can retrieve the operator via 
\bel{inverseTransform} 
\OpA = \int\ud\arbitraryParam \, \arbitraryCF{A}^{(s)}(\arbitraryParam)\inverseKernel(\arbitraryParam). 
\ee
Extending from \Eq{inverseTransform}, and following Ref.~\cite{Varilly:1989gs}, we can generate a generalized Fourier transform kernel to transform between any two phase-space functions with the same dimension by:
\be 
\arbitraryCF{A}^{(s_1)}(\arbitraryParam) = \int_{\arbitraryParam'} \ud \arbitraryParam' \, \arbitraryCF{A}^{(s_2)} (\arbitraryParam') \, \GenFourier(\arbitraryKernelN{1}(\arbitraryParam);\arbitraryKernelN{2}(\arbitraryParam'))
\ee
for
\be
\label{GenFourier}
\GenFourier(\arbitraryKernelN{1}(\arbitraryParam);\arbitraryKernelN{2}(\arbitraryParam')) \equiv \Trace{\arbitraryKernelN{1}(\arbitraryParam)\inverseKernelN{2}(\arbitraryParam')},
\ee
where the kernel on the right-hand side of the semicolon follows the inverse kernel from \Eq{inverseTransform}.
Using the two distinct subscripts on the kernel, $s_1$ and $s_2$, allows us to transform between any two phase-space functions, regardless of their respective ordering.
Following this, we can also express the trace of two operators as
\ba
\Trace{\OpA\OpB} = \iint  \ud\arbitraryParam \,&\ud\arbitraryParam'& \, \arbitraryCF{A}^{(s_1)}(\arbitraryParam) \arbitraryCF{B}^{(s_2)}(\arbitraryParam')  \label{funcTr}  \\
&\times& \Trace{\inverseKernelN{1}(\arbitraryParam)\inverseKernelN{2}(\arbitraryParam')}\nonumber. 
\ea
This can be extended to the trace of any number of operators, as long as the ordering of the kernels in the trace on the right hand side of the equation correspond to the same order of the the operators on the left side of the equation. 
We also note that the different $s_i$ values allow us to take the trace of two operators from any two phase-space functions.
Lastly, the Hamiltonian dynamics of the system follows from the von Neumann equation and is given by 
\ba
\PD{\arbitraryCF{\rho}^{(s)}(\arbitraryParam)}{t} &=&-\frac{i}{\hbar}\Trace{\COM{\OpH}{\DO}\arbitraryKernel(\arbitraryParam)} \\
&=&-\frac{i}{\hbar}\Trace{\COM{\arbitraryKernel(\arbitraryParam)}{\OpH}\DO}, \nonumber
\ea
for some Hamiltonian $\OpH$ and density operator $\DO$~\cite{Moyal1949}. 

By using~\Eq{inverseTransform}, the evolution equation above can be written entirely in phase space as
\ba
\PD{\arbitraryCF{\rho}^{(s)}(\arbitraryParam)}{t} = -\frac{i}{\hbar}\iint &\ud\arbitraryParam'& \ud\arbitraryParam'' \arbitraryCF{H}^{(s)}(\arbitraryParam') \arbitraryCF{\rho}^{(s)}(\arbitraryParam'')   \\ 
&\times& \Trace{\COM{\arbitraryKernel(\arbitraryParam)}{\inverseKernel(\arbitraryParam')}\inverseKernel(\arbitraryParam'')}\nonumber.
\ea
This motivates an extension of~\Eq{GenFourier} that allows us to perform a convolution of two functions, generating a Moyal star product kernel:
\ba
\GenConvolution(\arbitraryKernelN{1}(\arbitraryParam); \arbitraryKernelN{2}&&(\arbitraryParam'), \arbitraryKernelN{3}(\arbitraryParam'')) \equiv \nonumber \\
&&\Trace{\arbitraryKernelN{1}(\arbitraryParam) \inverseKernelN{2}(\arbitraryParam') \inverseKernelN{3}(\arbitraryParam'')},
\ea
so that, by setting $s_i=s$, we can define a generalization of the usual star product following similar arguments by Klimov~\cite{Klimov2002} according to 
\ba
&&\arbitraryCF{A}^{(s)}(\arbitraryParam) \star \arbitraryCF{B}^{(s)}(\arbitraryParam) \equiv   \\&& \nonumber
\iint \ud \arbitraryParam' \ud \arbitraryParam'' \,    
\arbitraryCF{A}^{(s)}(\arbitraryParam')\arbitraryCF{B}^{(s)}(\arbitraryParam'') \,
\GenConvolution(\arbitraryKernel(\arbitraryParam);\arbitraryKernel(\arbitraryParam'),\arbitraryKernel(\arbitraryParam'')). \label{GenMoyStar}
\ea
We can then use this definition to write the system's dynamics purely in terms of a Moyal bracket, 
\ba
\{\{\arbitraryCF{A}^{(s)}(\arbitraryParam),&\arbitraryCF{B}^{(s)} &(\arbitraryParam)\}\} \equiv  \\
 &\arbitraryCF{A}^{(s)}&(\arbitraryParam)\star \arbitraryCF{B}^{(s)}(\arbitraryParam)-\arbitraryCF{B}^{(s)}(\arbitraryParam)\star\arbitraryCF{A}^{(s)}(\arbitraryParam), \nonumber
\ea
in the familiar form of a generalized Liouville equation
\bel{dynamics}
\PD{\arbitraryCF{\rho}^{(s)}(\arbitraryParam)}{t} = -\frac{\ui}{\hbar} \{\{\arbitraryCF{H}^{(s)}(\arbitraryParam), \arbitraryCF{\rho}^{(s)} (\arbitraryParam)\}\},
\ee
which is now fully equivalent to the quantum von Neumann equation for the system.
We note that for Heisenberg-Weyl (HW) systems this reduces, in the limit $\hbar \rightarrow 0$, to 
\bel{pqdynamics}
\PD{\arbitraryCF{\rho}^{(0)}(\mathbf{q},\mathbf{p})}{t} =\{\arbitraryCF{H}^{(0)}(\mathbf{q},\mathbf{p}), \arbitraryCF{\rho}^{(0)} (\mathbf{q},\mathbf{p})\} 
\ee
where $\{\cdot,\cdot\}$ is the usual Poisson bracket. For the Wigner function of position and momentum, Moyal showed that in the classical limit the Wigner symbol becomes the same as its classical counterpart so that $\arbitraryCF{H}^{(0)}(\mathbf{q},\mathbf{p})=H(\mathbf{q},\mathbf{p})$ and $\arbitraryCF{\rho}^{(0)}(\mathbf{q},\mathbf{p})=\rho(\mathbf{q},\mathbf{p})$~\cite{Moyal1949}. So we see that in this ``classical'' limit we simply regain,
\bel{pqdynamics}
\PD{{\rho}(\mathbf{q},\mathbf{p})}{t} =\{{H}(\mathbf{q},\mathbf{p}), {\rho} (\mathbf{q},\mathbf{p})\}, 
\ee the standard Liouville equation of classical mechanics.

The phase-space framework we present above is completely general and, while its evaluation can be non-trivial for some systems, modern computational symbolic algebra should render phase-space methods for many quantum systems usable. 
Different problems are more efficiently solved in different representations, such as Heisenberg matrix mechanics vs Feynman path integrals.
 Phase-space methods may render more tractable certain classes of problem not readily solvable by other methods (see, for example,~\cite{PhysRevLett.80.4361}). 
Examples could well include open quantum systems and quantum chemistry.  
We note that a number of authors including Moyal and Groenewold have produced similar arguments to the above although the presentation has tended to be in a more system-specific form~\cite{Groenewold1946, Weyl1927, Klimov2002, Weizenecker2018}.

\subsection{The Wigner function}

As in classical mechanics, a quantum statistical theory would not be complete (or as powerful) without also possessing the characteristic function complement of the probability density function.
We now set out the procedure for generating the kernels for the two functions we will be primarily interested in discussing here. 
These are the two needed to generate the Wigner and Weyl functions that were discussed for the HW group case in Section~\ref{Intro}. 
Since we are only considering these two functions, the kernel is symmetrically ordered ($s=0$) and so we drop the $s$ subscript so that $\GenKer_0(\arbitraryParam)\equiv\GenKer(\arbitraryParam)$.

As shown in~\Eq{NormWignerHW}, the Wigner function kernel for position and momentum space is generated from a displaced parity operator. 
To generalize the Wigner function kernel we follow Ref.~\cite{1601.07772} and use notions of both a generalized parity $\hat\Pi$ operator and a generalized displacement or shift operator.
The latter is denoted by $\GenWeyKer(\GenWigPar)$, where we will take $\Theta\rightarrow\GenWigPar$ for the generalized Wigner function.
It should also be noted that we will take $\Theta\rightarrow\Weyling{\GenWigPar}$ for the parameterization of the generalized Weyl function to display the difference between the parameterization for the Wigner function and the dual parameterization for the Weyl function. 

The displacement operator, $\GenWeyKer(\GenWigPar)$, can be seen as a shift operator that translates the vacuum state of the system in consideration to a valid coherent state. 
It must therefore have the property~\cite{PhysRev.131.2766}
\bel{Disp} 
\GenWeyKer(\GenWigPar)\ket{0} = \ket{\GenWigPar}
\ee
where $\ket{0}$ is the vacuum state for an arbitrary system and $\ket{\GenWigPar}$ is the displaced vacuum or generalized coherent state. 
Next, the generalized parity $\hat\Pi$ is set by the Stratonovich-Weyl conditions~\cite{Stratonovich56}, (taken and adapted from Ref.~\cite{1601.07772}) given by: 
{\renewcommand{\labelenumi}{\footnotesize{\upshape{S-W.\theenumi}}}
\begin{enumerate}
\item\label{SW1} The mappings $W_{\OpA}(\Omega)=\Trace{\OpA \, \hat{\Pi}(\Omega)}$ and $\OpA = \int_{\Omega} W_{\OpA}(\Omega) \hat{\Pi}(\Omega) \ud \Omega$ exist and are informationally complete. Simply put, we can fully reconstruct $\OpA$ from $W_{\OpA}(\Omega)$ and vice versa~\footnote{For the inverse condition, depending on the parity $\hat{\Pi}(\Omega)$ used, an intermediate linear transform may be necessary.}. Note that $\ud\Omega$ here is a volume normalized differential element.
\item\label{SW2} $W_{\OpA}(\Omega)$ is always real valued (when $\OpA$ is Hermitian) which means that $\hat\Pi$ must be Hermitian. 
\item\label{SW3}  $W_{\OpA}(\Omega)$ is ``standardized'' so that the definite integral over all space $\int_{\Omega} W_{\OpA}(\Omega) \ud \Omega = \Trace{\OpA}$  exists and $\int_{\Omega} \hat{\Pi}(\Omega) \ud \Omega =\Bid$.
\item\label{SW4} Unique to Wigner functions, $W_{\OpA}(\Omega)$ is self-conjugate; the definite integral $\int_{\Omega} W_{\OpA'}(\Omega)W_{\OpA''}(\Omega) \ud \Omega= \Trace{\OpA' \OpA''} $ exists. 
This is a restriction of the usual Stratonovich-Weyl correspondence. 
\item\label{SW5} Covariance:
Mathematically, any Wigner function generated by ``rotated'' operators $\hat{\Pi}(\Omega^{\prime})$ (by some unitary transformation $V$) must be equivalent to ``rotated'' Wigner functions generated from the original operator ($\hat{\Pi}(\Omega^{\prime}) \equiv V \hat{\Pi}(\Omega) V^{\dagger}$) - \textit{i.\ e.\ }if $\OpA$ is invariant under global unitary operations then so is $W_{\OpA}(\Omega)$.
\end{enumerate}
}
We can therefore generate the general Wigner function by this kernel (or a tensor product of such kernels) by setting 
\be
\text{Wigner kernel: }\GenKer(\arbitraryParam) \rightarrow \GenWigKer(\GenWigPar) \equiv \GenWeyKer(\GenWigPar)\hat\Pi\GenWeyKer^\dag(\GenWigPar) 
\ee
over some  parameterization $\GenWigPar$. 
Therefore, from \Eq{SW1}, the Wigner function is given by
\bel{GenWigEq}
\GenWigF{A}(\GenWigPar) = \Trace{\GenArg \, \GenWigKer(\GenWigPar)}.
\ee
We note that for Wigner functions,~\Eq{funcTr} reduces to S-W.\ref{SW4}.

\subsection{The Weyl function}
Here we move from summarizing and modernizing past work to the central finding of this paper that enabled us to bring together the various elements of phase-space methods into a single coherent whole -- completing the Wigner, Weyl, and Moyal program of work and forming our central results.

When generalizing the Wigner function to any quantum system we used the notion of displaced parity as a starting point combined with the Stratonovich-Weyl correspondence to determine the exact form of the kernel. 
As with the Wigner function, a key constraint for the Weyl function is that the transform to phase space must be informationally complete. 
We further require that the transform be invertible to the original operator in its Hilbert space according to~\Eq{inverseTransform}. 
Using the same strategy for the Weyl function we propose that its generalization, $\GenWeyF{A}$, is then simply obtained by using a kernel in direct analogy with that for the usual Weyl function, which is the displacement operator defined in~\Eq{Disp} (or a tensor product of such kernels for an ensemble), that is  
\be
\text{Weyl kernel: }\GenKer(\arbitraryParam) \rightarrow \GenWeyKer(\GenWeyPar) 
\ee
over some suitably chosen dual parameterization $\GenWeyPar$. 
As we will discuss below and later in the work, the choice of parameterization -- and the associated displacement operator -- has been, in our view, the major obstacle preventing past attempts to generalize the Weyl function from being successful.
We note for a given system there is no one unique displacement operator, and care must be taken in choosing one that satisfies our constraints. 
In order to ensure the condition of invertiblity according to~\Eq{inverseTransform} is met we note that the phase spaces for the Wigner and Weyl functions need not be of the same dimension. 
While this may at first seem surprising we will provide in Section~\ref{TheOAMS} below a specific example and discussion clarifying how and why this is needed. 
It is worth noting that the definition of the Weyl function is given by the expectation value of the displacement operator while the Wigner function also needs the notion of parity. 
For this reason the Weyl function might be considered more fundamental.

Using an appropriate displacement operator the Weyl function is thus defined as:
\bel{GenWeyEq}
\GenWeyF{A}(\GenWeyPar) = \Trace{\GenArg \, \GenWeyKer(\GenWeyPar)}.
\ee
From~\Eq{GenWeyEq}, $\GenArg$ can be reconstructed using~\Eq{inverseTransform} according to
\be
\GenArg =  \int \ud \GenWeyPar \; \GenWeyF{A}(\GenWeyPar) \GenWeyKer^\dag(\GenWeyPar) \label{GenWeylInverse}
\ee
%where $\ud \GenWigPar$ and $\ud \GenWeyPar$ are volume normalized differential elements. 
where $\ud \GenWeyPar$ is a volume normalized differential element. 
Using \Eq{GenFourier}, it is therefore possible to transform between the Wigner and Weyl functions in terms of each other according to:
\ba
\GenWeyF{A}(\GenWeyPar) &=& \int \ud \GenWigPar \; \GenWigF{A}(\GenWigPar) \; \GenFourier^*(\GenWigKer(\GenWigPar); \GenWeyKer(\GenWeyPar)) \label{WeylasWignerFourier}, \\
\GenWigF{A}(\GenWigPar) &=& \int \ud \GenWeyPar \; \GenWeyF{A}(\GenWeyPar) \; \GenFourier(\GenWigKer(\GenWigPar); \GenWeyKer(\GenWeyPar)). \label{WignerasWeylFourier}
\ea

%-----------------------------------------------------------------------------
%-----------------------------------------------------------------------------
\section{Example Systems}\label{SectionThree}

%-----------------------------------------------------------------------------
%-----------------------------------------------------------------------------
\subsection{The Heisenberg-Weyl Group}\label{SHO}
The full standard formalism, as described in the introduction for Wigner (Weyl) functions, is recovered by the parameterization of position $q$ ($\tilde{q}$) and momentum $p$ ($\tilde{p}$) [or $\alpha$ and $\tilde{\alpha}$] and using the usual displacement and parity operators. This is a textbook system and is described in the introduction.

%-----------------------------------------------------------------------------
%-----------------------------------------------------------------------------
\subsection{\SU{2} and Orbital Angular Momentum States}\label{TheOAMS}
Considering the phase-space functions for \SU{2} angular momentum states, we start again with the generation of the displaced parity operator for the Wigner function.
When considering \SU{2} we need to replace the displacement operator with the notion of a rotation operator that rotates a spin vacuum state to an arbitrary spin coherent state.
The problem we face is that such an operator is not unique. 
One choice of operator is given by Arecchi~\cite{PhysRevA.6.2211} and expanded on by Perelomov in Ref.~\cite{0038-5670-20-9-R02}. 
This operator is the rotation operator defined in the subspace of degenerate eigenstates of $\hat{J}^2$:
\bel{ArecchiRotation}
\hat{R}(\xi)=\exp \left(\xi\hat{J}_{+} - \xi^{*}\hat{J}_{-}\right).
\ee
Here $\xi \equiv \theta e^{-\ui\phi} / 2$, where $\phi$ is the azimuthal angle, $\theta$ is the ordinate, and $\hat{J}_{\pm} = \hat{\mathsf{J}}_2^M(1)\, \pm \, \ui \hat{\mathsf{J}}_2^M(2)$, where $M\equiv2j$ ($j$ being the azimuthal quantum number and $M$, while strictly speaking redundant, is used to make clear the link between this work and the substantial body of existing group theory literature). 
We use $\hat{\mathsf{J}}_{2}^{M}(1)$, $\hat{\mathsf{J}}_{2}^{M}(2)$, and $\hat{\mathsf{J}}_{2}^{M}(3)$ instead of $\OpJx$, $\OpJy$, and $\OpJz$ respectively to take into account all possible $j$ values (these are the generators of the algebra $\{\hat{\mathsf{J}}_N^M\}$ that are defined in Appendix~\ref{AppA}).
There is a similarity in form between~\Eq{displacementOp} and~\Eq{ArecchiRotation} in that in the limit of high $j$,~\Eq{ArecchiRotation} tends towards the displacement operator of~\Eq{displacementOp}~\cite{PhysRevA.6.2211}.

In earlier work~\cite{1601.07772} we opted instead to use the \SU{2} rotation operator parameterised by the full Euler angles, such that
\ba 
&&\hat{U}_{2}^M(\phi,\theta,\Phi) = \label{Eulers}\\
&&\exp \left(\ui \hat{\mathsf{J}}_{2}^{M}(3)\phi\right) \exp \left(\ui \hat{\mathsf{J}}_{2}^{M}(2)\theta \right)\exp\left(\ui\hat{\mathsf{J}}_{2}^{M}(3)\Phi\right). \nonumber
\ea
The connection between~\Eq{ArecchiRotation} and~\Eq{Eulers} can be found by noting that 
\bel{EA}
\hat{R}(\phi,\theta) = \hat{U}_{2}^M(\phi,\theta,-\phi).
\ee

Next, to obtain the Wigner function kernel we need the generalized parity for spin-$j$ \SU{2}. 
The generalized parity can be expressed as a weighted sum of diagonal Hermitian operators, given by $\mathtt{J}_z$, of the Lie algebra of $\su{M + 1}$ in the fundamental representation (the spin-1/2 representation) calculated by the procedure in Appendix~\ref{AppA}:
\bel{SU2Parity}
\hat \Pi \rightarrow \hat \Pi^M_{2}  =  \sum_{l=0}^{M} \beta_2^M(l) \, \mathtt{J}_{z}([l+1]^2-1).
\ee
For simplicity we define $\mathtt{J}_{z}(0)\equiv \Bid_{M+1}$.
Equation~(\ref{SU2Parity}) gives the form of the generalized parity operator, displaying it as a weighted sum of the diagonal elements of the associated Lie algebra. Although we don't express this form in detail here, we show below a method to generate the generalized parity operator that is more in line with the existing literature on orbital angular momentum states \cite{PhysRevA.49.4101, Varilly:1989gs, PhysRevA.59.971, Klimov2002}.
This means that the kernel for the Wigner function is
\bel{SU2WigKer}
\GenWigKer(\GenWigPar) \rightarrow \hat \Pi^M_{2}(\phi,\theta) = \hat{U}_{2}^M(\phi,\theta,\Phi)\hat \Pi^M_{2} {\hat{U}_{2}^M}{^\dag}(\phi,\theta,\Phi)
\ee
where, because $\hat \Pi^M_{2}$ is diagonal and thus $\Phi$ makes no contribution due to the Baker–Campbell–Hausdorff condition, the parameterization of the phase space is just $(\phi,\theta)$ which is equivalent to that for the Bloch sphere~\cite{1601.07772}. 
We note that~\Eq{ArecchiRotation} also works as a valid rotation operator for orbital angular momentum Wigner functions, which can be seen by the relation in~\Eq{EA}, and that the parity is a diagonal matrix. 

Equation~(\ref{SU2Parity}) is the broad solution for the generalized parity, a special case of which was given in Ref.~\cite{1601.07772}, that is based on observations from Ref.~\cite{Rundle2016} for product states and from Ref.~\cite{Klimov055303} wherein a given spin-$j$ \SU{2} Wigner operator was defined as:
\ba
\label{eq:Klimov}
&&\hat{K}_M(\varphi, \vartheta) = 2 \sqrt{\frac{ \pi}{M+1}} \sum_{l = 0}^{M} \sum_{m = -l}^{l} Y_{lm}^{*}(\varphi, \vartheta) \hat{T}_{lm}^j, \nonumber \\
&&\hat{T}_{lm}^j = \sqrt{\frac{2l+1}{M+1}} \sum_{m',n= -j}^{j} C_{jm',lm}^{jn}\ket{j,n}\bra{j,m'}.
\ea
Here, $Y_{lm}^{*}$ are the conjugated spherical harmonics and $C_{jm',lm}^{jn}$ are Clebsch-Gordan coefficients that couple two representations of spin $j$ and $l$ to a total spin $j$. 
It can be easily shown that
\ba
\label{eq:WFNs-NEW}
\SpinPi^M_{2}  &\equiv &  \hat{K}_M(0,0) \nonumber\\
&=& \sum_{l=0}^{M} \frac{2l+1}{M+1} \sum_{n=-j}^{j} C^{jn}_{jn,l0}\ket{j,n}\bra{j,n}
\ea
linking our formalism to the multipole expansions found in other works~\cite{PhysRevA.59.971, PhysRevA.49.4101, Varilly:1989gs, Klimov2002}. 
Note that although both~\Eq{SU2Parity} and~\Eq{eq:WFNs-NEW} sum over the same number of elements, $\beta^M_2(l)$ is not necessarily equal to $(2l+1)/(M+1)$; for instance $\beta^M_2(0) = 1/(M+1)$, but for general $l$ $\beta^M_2(l)$ is a more complicated sum.

\begin{figure*}[!t]
%\vspace{+20pt}
\includegraphics[width=\linewidth, trim = {2cm, 10.4cm, 1.9cm, 1.8cm}, clip = true]{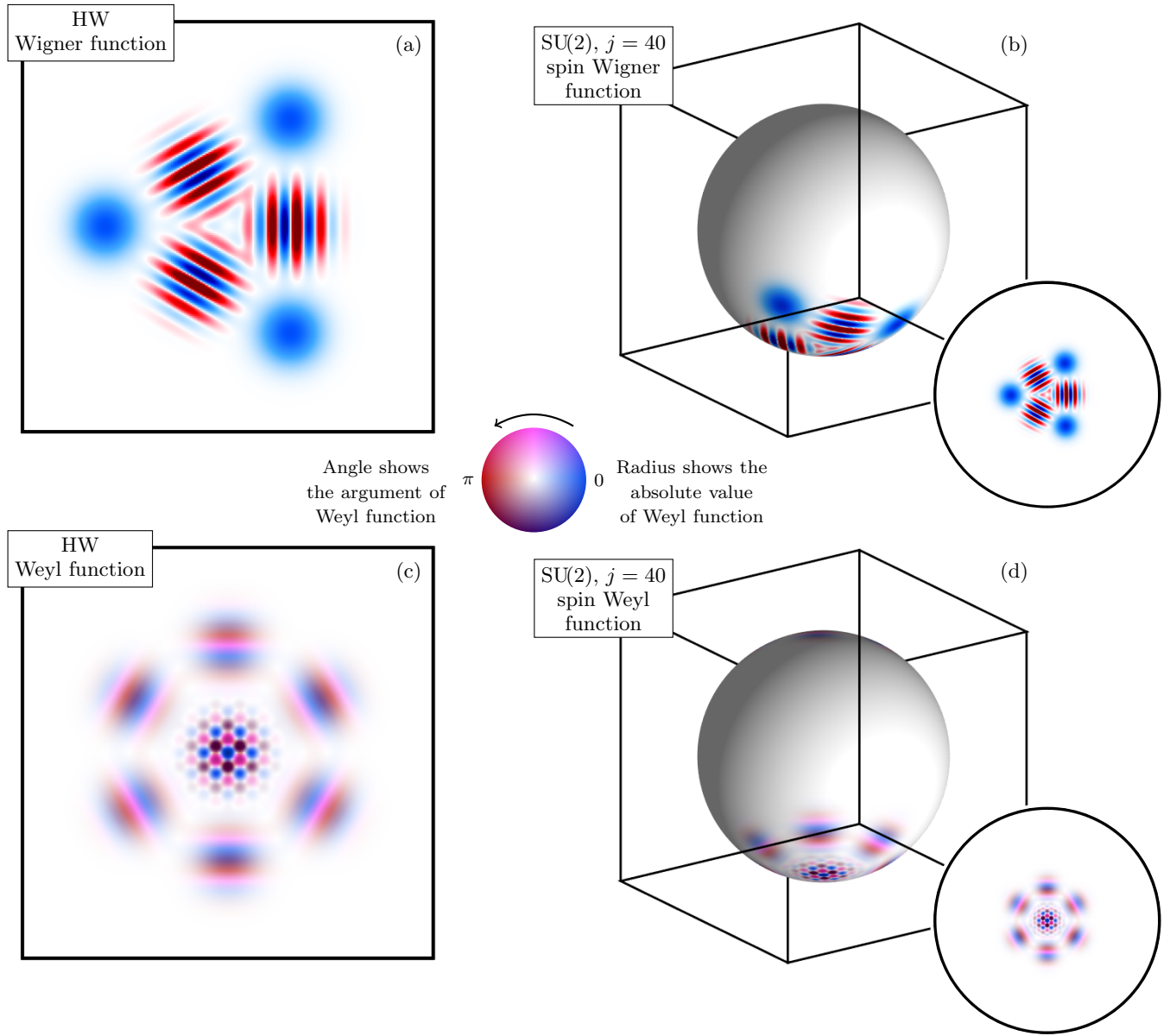}
%\vspace{+20pt}
\caption{\label{j40cats} Here we show (left column: a and c) the coherent superposition (Schr\"odinger cat) state of three macroscopically distinct coherent states: (a) is the Wigner function and (c) is the Weyl function. Each of the coherent states are generated from the displacement operator in \Eq{displacementOp}, such that $\ket{\alpha}=\OpD(\alpha)\ket{0}$. The state shown in (a) and (c) is explicitly $\ket{\psi} = \left(\ket{-3}+\ket{-3\exp(2\ui\pi/3)}+\ket{-3\exp(4\ui\pi/3)}\right)/\sqrt{3}$.
In the right column (b and d) we show a spin coherent state version of the state shown in the left column. 
These are a macroscopically distinct coherent superposition of spin coherent states (a spin Schr\"odinger cat) on the sphere where $j = 40$. 
Each of the ``cats" in this state has been created by applying the operator in~\Eq{SU2Weyl} to the lowest state $\ket{j;-j}$, such that a spin coherent state is given by $\ket{j;\phi,\theta} = \hat{R}(\phi,\theta)\ket{j;-j}$. The position of each spin coherent state with relation to the south pole is determined by the $\theta$ rotation.
Here $\theta = \pi/10$, as $j$ increases the value of $\theta$ will need to decrease to form the same analogue of a cat state seen in a continuous system, and thus in the stereographic projection, the spin coherent Schr\"odinger cat states at $\theta = \pi/10, \phi = \pi n/3$ ($n = 0,1,2$ for the three cats), will appear to get further away from each other. The state is explicitly given by $\ket{\psi} =\left(\ket{j;0,\pi/10}+\ket{j;\pi/3,\pi/10}+\ket{j;2\pi/3,\pi/10}\right)/\sqrt{3}$.
The inset next to each sphere in (b) and (d) is the corresponding stereographic (Riemann) projection of the lower hemisphere onto a circle in Euclidean space, with the boundary at the equator. 
Here (b) shows the spin Wigner function and (d) shows the spin Weyl function. 
Both (c) and (d) contain both magnitude (intensity) and phase (color) information for the complex valued Weyl functions as shown by the inset color wheel.}
\end{figure*}

\label{WeylPage}Unlike the Wigner function there have been few attempts to generate Weyl functions for spins. 
In our view, the most notable was proposed in Ref.~\cite{Haken1967} where the kernel is a rotation operator that is equivalent to the one defined in~\Eq{ArecchiRotation} (the equivalence to the operator used in Ref.~\cite{Haken1967} is shown in Ref.~\cite{PhysRevA.6.2211}).
The similarity of~\Eq{ArecchiRotation} and~\Eq{displacementOp} could lead one to believe that~\Eq{ArecchiRotation} would make a good kernel for the Weyl function given in~\Eq{GenWeyEq}. 
Unfortunately this kernel does not lead to a complete representation of the quantum state; the mapping from a density matrix to the Weyl function is not invertible by~\Eq{GenWeylInverse}. 
We therefore need to use \emph{instead} the rotation operator in~\Eq{Eulers} for our Weyl kernel to satisfy~\Eq{GenWeylInverse}:
\ba 
\label{SU2Weyl}
&&\GenWeyKer(\GenWeyPar)\rightarrow\hat{U}_{2}^M(\phiT,\thetaT,\PhiT) = \\ \nonumber
&&\exp \left(\ui \hat{\mathsf{J}}_{2}^{M}(3)\phiT\right) \exp \left(\ui \hat{\mathsf{J}}_{2}^{M}(2)\thetaT \right)\exp\left(\ui\hat{\mathsf{J}}_{2}^{M}(3)\PhiT\right).
\ea
For this reason the phase space of the Weyl function, having more degrees of freedom, is not the same as that of the Wigner function.
 Because the Weyl function is usually introduced as the two-dimensional Fourier transform of the Wigner function, this difference of phase space is why we asserted earlier in this work that the choice of parameterization and displacement operator formed the major obstacle in previous attempts to generalize the $(p-q)$ Weyl function to other systems. 
Although we use all three angles to define the Weyl function, when plotting we choose to use the slice from~\Eq{EA} where $\PhiT=-\phiT$ since this slice produces figures that are more in line with what is expected from a Weyl function (see~\Fig{j40cats} for an example).

For completeness, we note that Samson~\cite{Samson2000,Samson:2003ij}, and Scully and W\'odkiewicz~\cite{Scully1994}, made use of a similar characteristic function argument to generate Wigner functions with a phase space parametrized by three degrees of freedom. Their Wigner functions were generated by a kernel that was the Fourier transform of a characteristic function kernel. In both cases, this yielded a generalized delta function in place of Eq. (38). What is important to note is that in both of those works, the characteristic function was parameterized in terms of the symmetrized version of Tait–Bryan angles (pitch, roll, and yaw) rather than Euler angles. Consequently, in Ref.~\cite{Scully1994}, this formulation of the characteristic function was used to justify a delta function construction of the Wigner function. This lead to the problem that, although in \SU{2}, their Wigner functions, as a joint distribution of spin components, suffer from being singular. Our approach, on the other hand, overcomes all these issues by making use of the correct underlying quantum-mechanical group structure. Not only are all our distributions well behaved, this framework is also a more natural one since we interpret the Weyl function as the expectation value of a displacement operator and the Wigner function as the expectation value of a displaced parity operator.

%
%
% j = 40 spin cats

Due to the difference in degrees of freedom present in the functions, the volume normalized differential elements in~S.W-\ref{SW1} and~\Eq{GenWeylInverse} are not the same, this leads to the inverse transform to be given by
\be
\OpA = \int_{\Omega(\phi,\theta)}\ud\Omega(\phi,\theta)\,\GenWigF{A}(\phi,\theta) \,\Pi^M_{2}(\phi,\theta),
\ee
and
\be
\OpA = \int_{\Weyling{\Omega}(\phiT,\thetaT,\PhiT)}\ud\Weyling{\Omega}(\phiT,\thetaT,\PhiT)\,\GenWeyF{A}(\phiT,\thetaT,\PhiT)\,\hat{U}_{2}^{M\dagger}(\phiT,\thetaT,\PhiT),
\ee
where we can define the volume normalized differential elements to be 
\ba 
\ud \GenWigPar (\phi,\theta) &=&  \frac{M+1}{V_{\mathcal{C}\mathrm{P}^1}}\,\sin(2\theta)\,\ud\phi\,\ud\theta \label{WigVolNormD},\\
\ud\GenWeyPar(\phiT,\thetaT,\PhiT) &=& \frac{M+1}{V_{\SU{2}}} \, \sin(2\thetaT) \, \ud\phiT \, \ud\thetaT \, \ud\PhiT, \label{WeyVolNormD}
\ea
where the method to calculate $V_{\mathcal{C}\mathrm{P}^1}$ and $V_{\SU{2}}$ is shown in Appendix~\ref{AppC}.
In our view, the above differences in the phase-space structure for the Wigner and Weyl functions have been a major obstacle finding an invertible Weyl function for finite-dimensional systems. 
In this example the parametrization of the Weyl function is based on all three Euler angles $(\phiT,\thetaT,\PhiT)$. 
However, due to the parity being diagonal, the Wigner functions for \SU{2} appear to be parameterized by only two Euler angles $(\theta,\phi)$.

The fact still remains that both functions are parameterized over all three angles, although the diagonalization of the parity allows for the Wigner function to be defined on the manifold of pure states (\SU{N}/Z(N) -- where Z(N) is the center of \SU{N}) and the Weyl function exists in the full manifold (\SU{N}); that also means that either~\Eq{WigVolNormD} or~\Eq{WeyVolNormD} is an equally valid volume normalized differential element for the Wigner function. 
This therefore justifies the use of~\Eq{Eulers} as the best choice of rotation operator for both Wigner and Weyl functions.

%-----------------------------------------------------------------------------
%-----------------------------------------------------------------------------
\subsection{\SU{N}-symmetric Quantum Systems}\label{TheSUN}
The Wigner and Weyl functions for \SU{N} are found by generalizing the displacement and parity operators from the preceding section. 
Starting with the appropriate rotation operator,~\Eq{Eulers} has already conveniently been generalized to \SU{N} in Ref.~\cite{Tilma2}. 
The procedure to generate the \SU{N} rotation operators is shown in Appendix~\ref{AppB}.
The rotation operator is given by $\OpU_{N}^{M}({\boldsymbol{\phi}},{\boldsymbol{\theta}}, {\boldsymbol{\Phi}})$ for $\boldsymbol{\phi} = \{\phi_1 \ldots \phi_{N(N-1)/2} \}$, $\boldsymbol{\theta} = \{\theta_1 \ldots \theta_{N(N-1)/2} \}$, and $\boldsymbol{\Phi} = \{\Phi_1 \ldots \Phi_{N-1} \}.$

The parity is a straightforward generalization of~\Eq{SU2Parity} to~\SU{N}
\bel{NewParityGeneral} 
\hat \Pi \rightarrow \hat \Pi^M_N  = \sum_{l=0}^{d^M_N - 1} \beta_N^M(l) \, \mathtt{J}_{z}([l+1]^2-1)
\ee
where $d_{N}^{M}$ is the dimensionality of the system given by \Eq{TheOmega}. 
Here the $\mathtt{J}_z$ are the various diagonal hermitian operators of the Lie algebra of $\su{d_{N}^{M}}$ in the $M=1$ (i.e. fundamental) representation, as explained in detail in Appendix~\ref{AppA}.
The kernel for generating the Wigner function is therefore given by:
\bel{SUNWigner}
\GenWigKer(\SUNWigPar) \rightarrow \hat\Pi^M_{N}(\boldsymbol{\phi},{\boldsymbol{\theta}})= \OpU_{N}^{M}({\boldsymbol{\phi}},{\boldsymbol{\theta}}, {\boldsymbol{\Phi}}) \hat \Pi^M_{N}\OpU_{N}^{M}{^\dag}({\boldsymbol{\phi}},{\boldsymbol{\theta}}, {\boldsymbol{\Phi}}).
\ee
As with \SU{2} Wigner functions, the parity is diagonal which leads to the $\Phi_i$ terms canceling out. 
This in combination with further cancellations leaves the \SU{N} Wigner functions with $2(N-1)$ degrees of freedom, equally split between $\theta$ and $\phi$ degrees of freedom.
This split allows for the \SU{N} Wigner function to be visualized under an ``equal angle'' slicing that allows us to map the state to $\mathcal{S}_2$, allowing for a representation of \SU{N} in a generalized Bloch sphere similar to a Dicke state mapping~\footnote{For clarification, this mapping is not a Dicke state mapping; for instance, if we take the two options for a 3-level system mapping, \SU{3} and \SU{2} spin-1, these do not yield equivalent results on the 2-sphere.}. 

The explicit form of~\Eq{SUNWigner} for $M=1$ was given in Ref.~\cite{TilmaKae1} in terms of coherent states by
\ba 
&& \hat\Pi^1_{N}(\boldsymbol{\phi},{\boldsymbol{\theta}})= \frac{1}{N}\Bid_N + \label{Todd2012Wig}\\
&& \frac{\sqrt{N+1}}{2}\sum^{N^2-1}_{l=1}\bra{(\boldsymbol{\phi},\boldsymbol{\theta},\boldsymbol{\Phi})^1_N}\hat{\mathsf{J}}^1_N(l) \ket{(\boldsymbol{\phi},\boldsymbol{\theta},\boldsymbol{\Phi})^1_N}\hat{\mathsf{J}}^1_N(l), \nonumber
\ea
where $\hat{\mathsf{J}}^1_N(l)$ are the generalized Gell-Mann matrices given in Appendix~\ref{AppA}.
The coherent states in~\Eq{Todd2012Wig} are given by
\be 
\ket{(\boldsymbol{\phi},\boldsymbol{\theta},\boldsymbol{\Phi})^1_N} \equiv \OpU_{N}^{1}({\boldsymbol{\phi}},{\boldsymbol{\theta}}, {\boldsymbol{\Phi}})\ket{0},
\ee
where $\ket{0}$ is the lowest weighted (spin vacuum) state of dimension $d^1_N = N^2-1$~\cite{Nemoto2000}.
Using the same procedure used for~\Eq{eq:WFNs-NEW}, we can set $\boldsymbol{\theta} = \boldsymbol{\phi} = \boldsymbol{\Phi} = \mathbf{0}$ yielding the \SU{N} parity operator
\ba 
\hat\Pi^1_N &=& \frac{1}{N}\Bid_N + \frac{\sqrt{N+1}}{2} \bra{0}\hat{\mathsf{J}}_N^1(N^2-1) \ket{0}\hat{\mathsf{J}}_N^1(N^2-1)\nonumber \\
&=& \frac{1}{N}\left(\Bid_N - \sqrt{\frac{(N-1)N(N+1)}{2}}\hat{\mathsf{J}}_N^1(N^2-1)\right)
\ea
and returning the generalized parity operator given in Ref.~\cite{1601.07772}.

The kernel for generating the Weyl function is therefore also an extension of the \SU{2} case in~\Eq{SU2Weyl}, where we replace the \SU{2} rotation operator with the \SU{N} rotation operator used for the corresponding Wigner function in~\Eq{SUNWigner}, and so
\be
\GenWeyKer(\SUNWeyPar)\rightarrow\OpU_{N}^{M}(\Weyling{\boldsymbol{\phi}},\Weyling{\boldsymbol{\theta}}, \Weyling{\boldsymbol{\Phi}}). 
\ee 
We again note that this Weyl function has more degrees of freedom than the corresponding Wigner function. 
This is since the $N-1$ $\boldsymbol{\Phi}$ degrees of freedom make no contribution in the Wigner function but are still present in the Weyl function.
A comprehensive discussion can be found in~\cite{1601.07772}.

Given arbitrary \SU{N} Wigner and Weyl functions, $\GenWigF{A}(\GenWigPar)$ and $\GenWeyF{A}(\GenWeyPar)$, the density operators can be recovered again by using~S.W-\ref{SW1} and~\Eq{GenWeylInverse} respectively, where the normalized differential elements can be constructed using Appendix~\ref{AppC}.

%-----------------------------------------------------------------------------
%-----------------------------------------------------------------------------
\subsection{General Composite Quantum Systems}\label{TheGCQS}
Generalization to composite systems is, in principle, straightforward.
Consider a set of $\mathcal{N}$ quantum systems with respective Wigner and Weyl kernels being $\GenWigKer_i(\SUNWigPar_i)$ and $\GenWeyKer_i(\SUNWeyPar_i)$. 
Then the composite kernels for finding the total phase-space distributions are found simply by taking the tensor product of the respective kernels of each component system:
\ba
\GenWigKer(\CompWigPar)&\rightarrow&\bigotimes_i^\mathcal{N}\GenWigKer_i(\SUNWigPar_i),\\
\GenWeyKer(\CompWeyPar)&\rightarrow& \bigotimes_i^\mathcal{N} \GenWeyKer_i(\SUNWeyPar_i). 
\ea
Here, $\CompWigPar \rightarrow \{ \SUNWigPar_i,\ldots,\SUNWigPar_\mathcal{N} \}$ and $\CompWeyPar \rightarrow \{\SUNWeyPar_i,\ldots,\SUNWeyPar_\mathcal{N} \}$. 
The volume normalized differential elements to return the Hilbert space operator are therefore given by
\ba  
\ud \GenWigPar & \rightarrow & \prod_i^{\mathcal{N}} \ud \GenWigPar_i, \\
\ud \GenWeyPar & \rightarrow & \prod_i^{\mathcal{N}} \ud \GenWeyPar_i,
\ea
where the procedure to generate each of the $\ud\GenWigPar_i$ and $\ud\GenWeyPar_i$ is defined in Appendix~\ref{AppC}.

Following this scheme for the HW group returns the formalism for a collection of particles in position and momentum phase space $(\mathbf{q},\mathbf{p})$ as originally introduced by Wigner~\cite{Wigner1932}. 
Importantly, these kernels allow us to generate Wigner and Weyl functions for any composite system including hybrid ones (such as qubits and fields in  quantum information processing devices, atoms and molecules including both spatial and spin degrees of freedom, and particle physics in phase space). 
The fact that it is also possible to calculate quantum dynamics following~\Eq{dynamics} in phase space may lead to alternative pathways to numerical calculating a systems dynamics. 
For example an $N$ electron Wigner function, as might be applied in quantum chemistry, has $6N$ spatial and $2N$ spin continuous real degrees of freedom (rather than $3N$ complex continuous and $2^N$ discrete ones). 
It may be that such a representation could, in some situations, yield dynamics efficiently modeled by adaptive mesh solvers in regimes where traditional methods are not efficient (such as in modeling chemical reactions).
%
%%%%% j = 5/2 Superposition and 5-qubit GHZ state Figure %%%%%%
\begin{figure}[!t]
\includegraphics[width=\linewidth, trim = {1.9cm, 16.67cm, 10.7cm, 1.9cm}, clip = true]{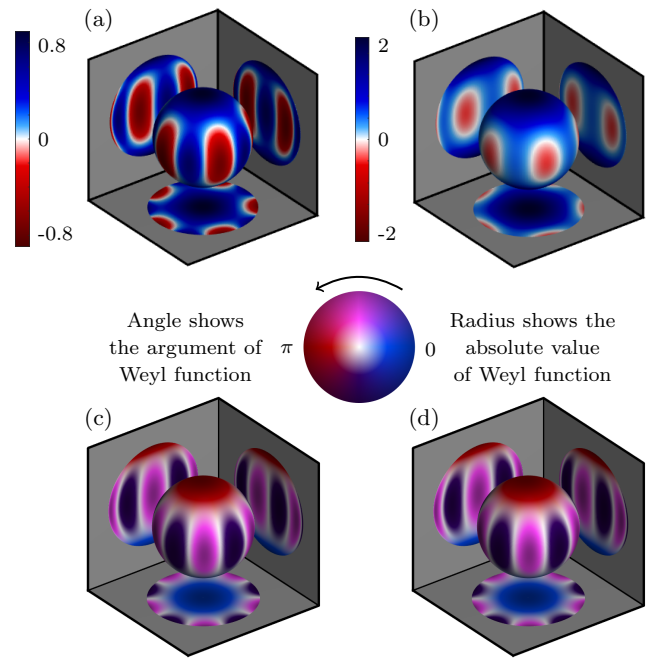}
\caption{\label{Cats} Here we show (a) and (c) the superposition state for a spin-$\frac{5}{2}$ spin coherent Schr\"odinger cat state~\cite{PhysRev.93.99}, given by $\left(\ket{-\frac{5}{2}}+\ket{\frac{5}{2}}\right)/\sqrt{2}$ and in (b)and (d) the five-qubit GHZ state $\left(\ket{00000}+\ket{11111}\right)/\sqrt{2}$.
Figures (a-b) show the spherical plot for the the spin Wigner functions where we have taken the equal angle slice, $\phi_i=\phi$ and $\theta_i=\theta$, and where blue is positive and red in negative; (c-d) give the spin Weyl functions spherical plots for the slice $\Weyling{\Phi}=-\Weyling{\phi}$, and where we have again taken the equal angle slice $\Weyling{\phi}_i = \Weyling{\phi}, \; \Weyling{\theta}_i = \Weyling{\theta}$. 
The phase for the spin Weyl functions is given by color according to the color wheel in the center of the figure. 
The absolute value is shown by saturation, so that the Weyl function is white when the value at that point is zero. 
Note that we have extended the range when mapping the function onto the sphere, so that the $\thetaT$ degree of freedom is doubled.
}
\end{figure}

Given $\mathcal{N}$ qudits, there are various ways a state can be shown in phase space.
Much of the previous work on Wigner functions for finite spaces have chosen a Dicke state~\cite{PhysRev.93.99} mapping of $\mathcal{N}$ qubits to an \SU{2} $M=\mathcal{N}$ function. 
In our earlier work~\cite{Rundle2016}, we chose to take either the tensor product of $\mathcal{N}$ \SU{2} kernels, 
\bel{CompKernel5}
\hat{\Pi}(\CompWigPar) = \otimes^{\mathcal{N}}_{i=1}\hat{\Pi}^1_2(\Omega_i),
\ee
or to take $\mathcal{N}$ \SU{2} rotation operators, $\OpU(\CompWigPar) = \otimes^{\mathcal{N}}_{i=1}\OpU^1_2(\Omega_i)$, with the \SU{2^\mathcal{N}} parity.
As an example, in~\Fig{Cats} we compare two of the options for visualizing a 5-qubit GHZ state. 
In the first column, (a) and (c), we show the Wigner and Weyl function according to Section~\ref{TheOAMS}, where $M = 5$. 
This state can be interpreted as either the approximation of the 5 qubit GHZ state or a 6-level angular moment state in a superposition of the highest and lowest weighted state. 
In the second column, (b) and (d), we show the 5 qubit GHZ state with a tensor product of 5 \SU{2} kernel shown in~\Eq{CompKernel5} for the Wigner function and the tensor product of the rotation operator for the Weyl function. 
Since for these visualizations we have $10$ ($15$ for the Weyl function) degrees of freedom, unlike the $2$ ($3$ for Weyl) degrees of freedom needed for the Dicke states, we need to choose appropriate slices. 
For the Wigner function we have taken the equal angle slice $\theta_i = \theta$ and $\phi_i = \phi$. 
For the Weyl function we have set $\thetaT_i = \thetaT$, $\phiT_i = \phiT$, and $\PhiT_i=-\phiT$. 

We can see from \Fig{Cats} that the two Wigner functions (a-b) look similar, this is since the equal angle slice is similar to the symmetric subspace. 
Although the two Wigner functions look similar, the advantage of using the tensor product state can be found in the fact it is informationally complete, whereas a Dicke state mapping is not.
Interestingly, the Weyl function for the two different choices of kernel are identical. 
The Wigner functions differ due to the weighting given by the parity to each element of the given basis, since the parity isn't present in the Weyl kernel such a weighting doesn't exist and every element is equally weighted.

%-----------------------------------------------------------------------------
%-----------------------------------------------------------------------------
\section{Quantum Statistical Mechanics in Phase Space}\label{enthalpy}

Both the Weyl formalism developed here and the Wigner formalism given in~\cite{1601.07772,Rundle2016} allow us to analyze finite-dimensional and composite quantum systems in the same way as one would analyze continuous-variable quantum systems.
Both the Wigner and Weyl functions are informationally complete; one can always regain the Hilbert space representation of the collection of states by suitable integration of the parameters for the phase-space functions with the appropriate kernel. 
A corollary to this condition is that any quantum-mechanical property defined in Hilbert space must have an equivalent phase space definition.
The close relationship between quantum phase-space methods as presented here and other statistical methods is apparent from~\Eq{dynamics}, which takes the form of a generalized Liouville equation.
Furthermore, as one can now discuss and define thermodynamic concepts and quantities for collections of finite quantum systems~\cite{1751-8121-49-14-143001,Brando17032015}, it goes without saying that one can have the same discussion by using the Weyl or Wigner function of the same collection of states.

This connection is well know to be more than a superficial one. 
For instance, the partition function $Z(\beta)$ can be found
following the same approach as originally suggested by Wigner~\cite{Wigner1932}. 
For a given unnormalized thermal density matrix $\hat{\rho}(\beta) = \exp(-\beta \OpH)$ where $\beta \equiv 1/k_B T$ 
\bel{Partition} 
Z(\beta) \equiv \Trace{\hat{\rho}(\beta)}=\int_{\Omega} \ud \Omega \, W_{\hat{\rho}(\beta)}(\Omega),
\ee
making use of S.W-\ref{SW3}. Interestingly, to first order in $\beta$ we see a direct connection between the Wigner function for the Hamiltonian $W_{\OpH}(\Omega)$ and the partition function
\ba 
Z(\beta)
   &=&\int_{\Omega} \ud \Omega \, W_{\hat{\rho}(\beta)}(\Omega) \nonumber \\
   &=&\int_{\Omega} \ud \Omega \, \Trace{\hat{\rho}(\beta)\hat{\Pi}(\Omega)} \nonumber \\
   &=&\int_{\Omega} \ud \Omega \, \Trace{\sum^{\infty}_{n=0} \frac{(-\beta \OpH)^n}{n!}  \hat{\Pi}(\Omega)} \nonumber \\
   &=& Z(0) - \beta \int_{\Omega} \ud \Omega \, W_{\OpH}(\Omega) \nonumber \\
   && \hspace{0.8cm} + \frac{\beta^2}{2} \int_{\Omega} \ud \Omega \, W_{\OpH}^2(\Omega) + \mathcal{O}(\beta^3),
\ea
where the second and third terms are easily calculated and come directly from S.W-\ref{SW3} and S.W-\ref{SW4} respectively.
It also follows from S.W-\ref{SW3} that $Z(0)$ is the dimensionality of the Hilbert space.
We note that for some systems there may be a computational advantage to using the above approach to compute the approximate partition function, in particular for small values of $\beta$.
From the partition function we can further calculate other thermodynamical quantities such as the total energy
\be 
\langle E\rangle = -\frac{\partial \ln Z(\beta)}{\partial \beta},
\ee
and free energy
\be 
A = -\frac{1}{\beta} \ln Z(\beta),
\ee
with clear analogy to classical statistical mechanics. This will be of utility in the burgeoning field of quantum thermodynamics.

When using these methods to generate partition functions for finite systems, there are interesting cases for the expansion of~\Eq{Partition}. 
As an example, we consider the Pauli matrices in \SU{2}, given by $\mathbf{h}\cdot\hat{\boldsymbol{\sigma}}$, where $\mathbf{h}=[h_x,h_y,h_z]$ is the magnetic field. 
Setting $\hat{\rho}(\beta) = \exp(-\beta \mathbf{h}\cdot\hat{\boldsymbol{\sigma}})$, \Eq{Partition} reduces to
\ba 
Z(\beta) &=& \int_\Omega \ud \Omega \, \bigg( \cosh(\beta|\mathbf{h}|) \, W_{\Bid}(\Omega) \nonumber\\
& & \hspace{1.5cm} - \frac{\sinh(\beta|\mathbf{h}|)}{|\mathbf{h}|}\,W_{\mathbf{h}\cdot\hat{\boldsymbol{\sigma}}}(\Omega) \bigg) \nonumber \\
&=& 2\cosh(\beta|\mathbf{h}|). \label{PauliPartition}
\ea
It's useful to note that 
\be 
W_{\mathbf{h}\cdot\hat{\boldsymbol{\sigma}}}(\Omega) = h_xW_{\sigma_x}(\Omega)+h_yW_{\sigma_y}(\Omega)+h_zW_{\sigma_z}(\Omega)
\ee
which allows us to calculate the partition function through the Wigner functions of the individual Pauli matrices.
Furthermore, the mean value, $\bar{A}$, of any physical quantity, $\OpA$, is $\Trace{\OpA \exp(-\beta\OpH)}/Z(\beta)$. We note that this can be written (by using S.W-\ref{SW4}) in terms of the Wigner functions as
\bel{MeanVal}
\bar{A}  = \frac{1}{Z(\beta)}\int_\Omega \ud\Omega\, W_{\OpA}(\Omega)W_{\hat{\rho}(\beta)}(\Omega).
\ee 
By using the first line of~\Eq{PauliPartition}, we can extend this with~\Eq{MeanVal} to yield the solution
\be
\bar{A} = \frac{1}{2} \int_\Omega \ud \Omega  \, \bigg( W_{\OpA}(\Omega) - \frac{\tanh(\beta|\mathbf{h}|)}{|\mathbf{h}|} \, W_{\OpA}(\Omega)W_{\mathbf{h}\cdot\hat{\boldsymbol{\sigma}}}(\Omega)\bigg).
\ee
setting $A = \mathbf{e}\cdot\hat{\boldsymbol{\sigma}}$, for $\mathbf{e}=[e_x, e_y, e_z]$ where each $e_i$ is the component of magnetization in the $i$ direction, and noting that
\ba 
\int_\Omega &\ud \Omega & \,W_{\mathbf{e}\cdot\hat{\boldsymbol{\sigma}}}(\Omega) W_{\mathbf{h}\cdot\hat{\boldsymbol{\sigma}}}(\Omega)  \\
&=& \int_\Omega \ud \Omega \left( e_x h_x W_x^2(\Omega)+e_y h_y W_y^2(\Omega)+e_z h_z W_z^2(\Omega)\right),\nonumber
\ea
\Eq{MeanVal} reduces to the expected
\ba 
\bar{A} &=& -\frac{\tanh(\beta|\mathbf{h}|)}{|\mathbf{h}|}\,\mathbf{e}\cdot\mathbf{h}\nonumber \\
&=& -|\mathbf{e}| \cos (\vartheta) \tanh (\beta|\mathbf{h}|),
\ea
where $\vartheta$ is the angle between $\mathbf{e}$ and $\mathbf{h}$. So
$\bar{A}$ is therefore completely calculable with the Wigner function.

We now turn our attention to the Weyl function. 
%The first and simplest property of the Weyl function is that we can obtain the autocorrelation function for any state/observable with respect to any degree of freedom by fixing all other degrees of freedom apart from the one of interest to be zero~\cite{Chountasis1997}. 
%More importantly though is that it 
The Weyl function can be viewed as a quantum analog of the characteristic function~\cite{Moyal1949}.
In classical probability theory the Fourier transform of the probability density function is the characteristic function that has the powerful property of being a moment-generating function.
By following Refs~\cite{Wigner1984, Carmichael, PerelomovB} we can see that the Weyl function can be considered the quantum analog of this characteristic function. 
In particular, we see it acts as a moment generating function if we  consider some operator $\OpA$ where the phase space is parameterized by $\SUNWeyPar = \{\Weyling{\omega}_1, \ldots, \Weyling{\omega}_n\}$ where each $\Weyling{\omega}_i$ is an individual degree of freedom, so that each moment is 
\bel{MGF} 
M_{m_1,\ldots,m_n} = \prod_{i=1}^{n} \left(\eta_i\frac{\partial}{\partial \Weyling{\omega}_i}\right)^{m_i} \;\GenWeyF{A}(\GenWeyPar)|_{\SUNWeyPar = 0}
\ee
where $\eta_i=\pm1,\pm\ui$ depending on the sign in front of the corresponding moment in the generalized displacement operator.
For example, when looking at \SU{2} systems, to get the correct sign $\eta_i = -\ui$. For HW, when choosing moments of $\alpha$ ($\alpha^*$) the correct value is $\eta_i = -1$ (or just $1$ for $\alpha^*$).

Weyl or Wigner functions can be used in  in the generation of correlation functions. 
Correlation functions can be defined either in terms of time or spatial coordinates and in special cases can be rewritten as autocorrelation functions.
For example, the ambiguity function is the signal processing analog of the Weyl function that can be reduced to a temporal autocorrelation function by noting the spatial coordinates where the Doppler shift is zero. 
Similarly, when looking at the Weyl function from \Eq{WeylCompact}, by setting either $\Weyling{p}=0$ ($\Weyling{q}=0$) we can generate the autocorrelation function for position (momentum)~\cite{Chountasis1997}. 
This can be seen from the definition of a general autocorrelation function:
\be
{R}(\chi) = \int^\infty_{-\infty}\psi^*(s+\chi)\psi(s) \, \ud s 
 = \Trace{\hat \rho \,\OpD (\chi) }.
\label{TheACDef}
\ee
By extension to finite-dimensional systems is now possible by direct analogy. For example when considering a single spin we can define the following autocorrelation functions 
\be
\tilde{W}_{\hat\rho}(0,\thetaT,0) = \mathrm{Tr}\left[\hat \rho\; \hat U (0,\thetaT,0) \right] \equiv {R}(\thetaT) \label{spinAutocorrelationTheta}
\ee
and
\be
\tilde{W}_{\hat\rho}(\phiT,0,0) = \mathrm{Tr}\left[\hat \rho\; \tilde U (\phiT,0,0) \right] \equiv R(\phiT). \label{spinAutocorrelationPhi}
\ee
If we evaluate $R(\PhiT)$ we see that it is identical to \Eq{spinAutocorrelationPhi}; this allows us to view standard Weyl functions as \textit{effective} autocorrelation functions in the ``rotation and phase'' spin degrees of freedom.
Generalization of autocorrelation to any system is then simply given by
\bel{PhaseRFinal}
R(\Weyling{\omega}) = \Trace{\hat\rho\,{\hat{\mathcal{D}}}(\Weyling{\omega})},
\ee
where $\Weyling{\omega}$ is any degree of freedom from the full parameterization.
As the Weyl function is a characteristic function this relation to auto-correlation is expected.

Higher order correlation functions can be generated from directly measuring the Wigner or Weyl function by evaluating the continuous cross-correlation integral of the Wigner (Weyl) function with itself at a later time (corresponding to the mapping ${\Omega} \mapsto {\Omega} + {\mho}$ ($\Weyling{\Omega} \mapsto \Weyling{\Omega} + \Weyling{\mho}$), where %${\omega} = \{\boldsymbol{\nu}, \boldsymbol{\mu}, \boldsymbol{\upsilon}\}$) 
$\mho$ ($\Weyling{\mho}$) is the displacement in phase space, which yields:
\ba
 \mathcal{R}(\mho) &=& \frac{1}{\mathrm{V}_{\Omega}}\int_{\Omega} \ud \Omega \, W_{\hat\rho}(\Omega+\mho) W_{\hat\rho}(\Omega), \nonumber \\
\Weyling{\mathcal{R}}(\Weyling{\mho}) &=& \frac{1}{\mathrm{V}_{\Weyling{\Omega}}}\int_{\Weyling{\Omega}} \ud \Weyling{\Omega} \, \tilde{W}_{\hat\rho}(\Weyling{\Omega}+\Weyling{\mho}) \tilde{W}^{*}_{\hat\rho}(\Weyling{\Omega}). \label{GeneralRAC}
\ea
These are alternative forms of \Eq{PhaseRFinal}, in particular \Eq{spinAutocorrelationTheta} and \Eq{spinAutocorrelationPhi}, for the Wigner or Weyl function.
Following the discussion in Section~\ref{TheGCQS}, the extension of \Eq{GeneralRAC} to collections of systems, and thus comparisons to \Eq{PhaseRFinal}, is straightforward.

The Wiener–Khinchin theorem allows us to relate the autocorrelation functions defined in \Eq{GeneralRAC} to appropriate power spectral density functions (such as those used in neutron scattering~\cite{PhysRevB.47.11788}), via a Fourier transform.
More generally, it is clear that one can define a correlation function $\mathfrak{C}$ of a Weyl function of a collection of finite quantum systems at time $t_1$ and $t_2$, where $t_1 > t_2$, as 
\bel{correlationfunction2}
\mathfrak{C}(\tilde{\Omega}_1,\tilde{\Omega}_2) = \langle \tilde{W}_{\rho(t_1)}(\tilde{\Omega}_1) \, , \, \tilde{W}_{\rho(t_2)}(\tilde{\Omega}_2) \rangle,
\ee
and that the corresponding Wigner function version is generatable by exploiting \Eq{WignerasWeylFourier}.
What is more powerful is that we can define not two-point correlation functions, but $n$-point correlation functions of phase space functions:
\begin{multline}
\label{correlationfunctionN}
\mathfrak{C}(\tilde{\Omega}_1,\tilde{\Omega}_2, \ldots, \tilde{\Omega}_n) = \langle \tilde{W}_{\rho(t_1)}(\tilde{\Omega}_1) \, , \, \tilde{W}_{\rho(t_2)}(\tilde{\Omega}_2) \\ 
\, ,\ldots, \, \tilde{W}_{\rho(t_{n-1})}(\tilde{\Omega}_{n-1}) , \, \tilde{W}_{\rho(t_n)}(\tilde{\Omega}_n) \rangle.
\end{multline}
In this way, we map the changes in physical position and time to changes in phase-space coordinates, allowing us to define highly generalized static and dynamic structure factors for spin systems.

We believe that these ideas can be further applied to quantum statistical mechanics by using the above notions in lieu of the moments of the Inverse Participation Ratio (IPR)~\cite{0034-4885-56-12-001} in order to describe the localization and complexity of a collection of qubits or other quantum states, in particular those used in Anderson localization~\cite{AndersonLocalization}.
This will be the subject of future work.

%-----------------------------------------------------------------------------
%-----------------------------------------------------------------------------
\section{Conclusions}

In this work we have completed the Wigner-Weyl-Moyal-Groenewold program of work describing quantum mechanics as a statistical theory~\cite{Moyal1949, Groenewold1946}. 
We have presented the general framework in a modern context. 
Importantly we have shown how unifying concepts of displacement and parity lead to generalizations of Wigner and Weyl functions for any quantum system and its dynamics.  
For correctly formulating the Weyl function of a system we have discussed how taking proper account of its underlying group structure is essential. 
Specifically we observe that the Weyl function is not simply the two-dimensional Fourier transform of the Wigner function but is instead defined though a specific displacement operator and its parameterization. 
The fact that the dimensionalities of the two phase spaces differ has, in our view, been the major obstacle to completing the description of quantum mechanics as a statistical theory in phase space which we have here overcome.  
We have shown how a generalization of the Fourier transform links these two representations. 

We have also shown how we can utilize phase space to gain insight into statistical properties of quantum systems. 
We have shown how statistically important quantities such as the partition function and moment generating function can be constructed within this quantum phase space approach. 
This should lead to a natural framework for the study of important applications in fields such as quantum thermodynamics.

We speculate that, because we utilize only the underlying group structure of the system of interest, extensions to this work in areas outside of quantum mechanics may provide new insights. 
Of particular interest would be applications to signal processing where Wigner and Weyl (ambiguity) functions already find great utility. 
There have already been attempts to describe signal processing in terms of group actions (such as Ref.~\cite{Howard2006} and Ref.~\cite{Stergioulas2004}); a complete formalism could lead to more computational efficiency in many areas of the field. 
We might also borrow ideas from signal processing and ambiguity functions, such as the formulation of the energy, $E_{f}$, of a signal~\cite{ricker2003echo}.
Lastly, phase-space methods have seen many uses as entropic measure, such as the R\'enyi entropy~\cite{Renyi1961}; its extensions~\cite{PhysRevA.51.2575} link ideas in quantum and classical information theory. 

Finally it has been shown that by making use of its underlying group structure we can fully describe any quantum system in terms of a statistical theory in phase space. 
Because of this, not only is this theory capable of describing and providing new insights into standard quantum systems such as qubits, atoms, and molecules but we also propose that extensions to this would be of utility for systems with more exotic group structures such as E(8), \SU{1,1}, and anti-de Sitter space calculations.

%-
\acknowledgments
The authors would like to thank Kae Nemoto, Andrew Archer, A. Balanov and Luis G. MacDowell for interesting and informative discussions. 
TT notes that this work was supported in part by JSPS KAKENHI (C) Grant Number JP17K05569. 
RPR is funded by the EPSRC [grant number EP/N509516/1]. 
%------------------------------------------------------------------------

%\clearpage
\appendix
\begin{widetext}

\section{Generalized Pauli Matrices}\label{AppA}

The $\hat{\mathsf{J}}_{N}^{M}(k)$ are generalized Pauli matrices of dimension $d_{N}^{M}$ that are generated in the following way~\cite{Nemoto2000}:
\begin{enumerate}
\item{Define a general basis $\ket{m_1,m_2,\ldots,m_N}$ where $M=\sum_{k=1}^{N}m_k$, $m_{k} \in \mathbb{Z}$, and $2j \equiv M \in \mathbb{Z^{+}}$.}
\item{Define the following operators:
\begin{equation*}
\label{General-Lambda-Jab}
J_b^a \ket{m_1,\ldots,m_a, \ldots,m_b,\ldots,m_N} = \sqrt{(m_a+1)m_b} \, \ket{m_1,\ldots,m_a+1,\ldots,m_b-1,\ldots,m_N} 
\end{equation*}
for $1 \leq a < b \leq N$,
\begin{equation*}
\label{General-Lambda-Jba}
J_b^a \ket{m_1,\ldots,m_b,\ldots,m_a,\ldots,m_N} = \sqrt{m_a(m_b+1)} \, \ket{m_1,\ldots,m_a-1,\ldots,m_b+1,\ldots,m_N} 
\end{equation*}
for $1 \leq b < a \leq N$, and,
\begin{equation*}
\label{General-Lambda-C}
J_c^c \ket{m_1,\ldots,m_c,\ldots,m_N} = \sqrt{\frac{2}{c(c+1)}} \, \left(\sum_{k=1}^c m_k-c \, m_{c+1}\right)\ket{m_1,\ldots,m_c,\ldots,m_N} \nonumber
\end{equation*}
for $1 \leq c \leq N-1$.
}
\item{Using the basis given in 1 and the operators given in 2, define the following matrices:
\begin{eqnarray}
\label{General-Lambda-Final}
\mathtt{J}_{x}(a,b) &\equiv& J_b^a + J_a^b, \nonumber \\ 
\mathtt{J}_{y}(a,b)&\equiv& -\ui \left(J_b^a - J_a^b \right), \nonumber \\ 
\mathtt{J}_{z}([c+1]^2-1)&\equiv& J_c^c. 
\end{eqnarray}
for $a,b = 1,2,3,\ldots,N;\, a < b$ and $c=1,2,\ldots,N-1$.
}
\item{Combine the three matrices given in~\Eq{General-Lambda-Final} to yield the set $\{ \hat{\mathsf{J}}_{N}^{M}(k) \}$ where $k=1,2,\ldots,N^2-1$ and 
\bel{LRules-G2}
\Tr{\left[\hat{\mathsf{J}}_{N}^{M}(i) \cdot \hat{\mathsf{J}}_{N}^{M}(j)\right]} = \frac{2M}{N+1} d_{N+1}^{M}\delta_{ij},
\ee
}
\end{enumerate}
where 
\bel{TheOmega}
d_{N}^{M} \equiv \frac{(N+M-1)!}{M!(N-1)!}.
\ee
For example, for $N = 3$ and $M = 1$,~\Eq{General-Lambda-Final} gives the following $8$ matrices, the spin-$1/2$ \SU{3} hermitian operators also known as the Gell-Mann matrices~\cite{Greiner}:
\bel{SU3-lambdas}
\begin{array}{crcr}
&\mathtt{J}_{x}(1,2) \equiv \hat{\mathsf{J}}_{3}^{1}(1) = \left( \begin{array}{cccc}
                     0 & 1 & 0 \\
                     1 & 0 & 0 \\
                     0 & 0 & 0 \end{array} \right), \qquad
&\mathtt{J}_{y}(1,2) \equiv \hat{\mathsf{J}}_{3}^{1}(2) = \left( \begin{array}{crcr} 
                     0 & -\ui & 0 \\
                    \ui &  0 & 0 \\
                     0 &  0 & 0 \end{array} \right), \\
&\mathtt{J}_{x}(1,3) \equiv \hat{\mathsf{J}}_{3}^{1}(4) = \left( \begin{array}{crcr} 
                     0 & 0 & 1 \\
                     0 & 0 & 0 \\
                     1 & 0 & 0 \end{array} \right), \qquad
&\mathtt{J}_{y}(1,3) \equiv \hat{\mathsf{J}}_{3}^{1}(5)= \left( \begin{array}{crcr} 
                     0 & 0 & -\ui \\
                     0 & 0 &  0 \\
                     \ui & 0 &  0 \end{array} \right), \\
&\mathtt{J}_{x}(2,3) \equiv \hat{\mathsf{J}}_{3}^{1}(6)= \left( \begin{array}{crcr} 
                     0 & 0 & 0 \\
                     0 & 0 & 1 \\
                     0 & 1 & 0 \end{array} \right), \qquad
&\mathtt{J}_{y}(2,3) \equiv \hat{\mathsf{J}}_{3}^{1}(7) = \left( \begin{array}{crcr} 
                     0 & 0 &  0 \\
                     0 & 0 & -\ui \\
                     0 & \ui &  0  \end{array} \right),
\end{array}    
\ee
and
\bel{SU3-CSA}
\begin{array}{crcr}
&\mathtt{J}_{z}([1+1]^2-1) \equiv \hat{\mathsf{J}}_{3}^{1}(3)= \left( \begin{array}{crcr} 
                     1 &  0 & 0 \\
                     0 & -1 & 0 \\
                     0 &  0 & 0 \end{array} \right), \qquad
&\mathtt{J}_{z}([2+1]^2-1) \equiv \hat{\mathsf{J}}_{3}^{1}(8) = \dfrac{1}{\sqrt{3}}\left( \begin{array}{crcr} 
                     1 & 0 &  0 \\
                     0 & 1 &  0 \\
                     0 & 0 & -2 \end{array} \right).
\end{array}
\ee
Similarly, for for $N = 2$ and $M = 2$,~\Eq{General-Lambda-Final} gives the following spin-$1$ \SU{2} hermitian operators:
\be
\begin{array}{crcr}
&\mathtt{J}_{x}(1,2) \equiv \hat{\mathsf{J}}_{2}^{2}(1) = \dfrac{1}{\sqrt{2}}\left(\begin{array}{crcr}
                    0 & 1 & 0 \\
                    1 & 0 & 1 \\
                    0 & 1 & 0 \end{array} \right), \qquad
&\mathtt{J}_{y}(1,2) \equiv \hat{\mathsf{J}}_{2}^{2}(2) = \dfrac{1}{\sqrt{2}}\left(\begin{array}{crcr}
                   0 & -\ui & 0 \\
                   \ui & 0 & -\ui \\
                   0 & \ui & 0 \end{array} \right)
\end{array}    
\ee
and
\be
\begin{array}{crcr}
&\mathtt{J}_{z}([1+1]^2-1) \equiv \hat{\mathsf{J}}_{2}^{2}(3)= \left( \begin{array}{crcr}
                     1 &  0 & 0 \\
                     0 &  0 & 0 \\
                     0 &  0 & -1 \end{array} \right).
\end{array}
\ee
For completeness, we define $\mathtt{J}_{z}(0) \equiv \Bid_{d_{N}^{M}}$.

%-----------------------------------------------------------------------------
%-----------------------------------------------------------------------------
\section{$\hat{U}_{N}^{M}$ Operators}\label{AppB}

Our Weyl and Wigner formulations are based on the exploitation of a \SU{N} group action $\hat{U}_{N}^{M}$ given in~\cite{Tilma2}: 
\bel{eq:suN} 
\hat{U}_{N}^{M}(\boldsymbol{\phi},\boldsymbol{\theta}, \boldsymbol{\Phi}) = \biggl(\prod_{N \geq q \geq 2} \, \prod_{2 \leq p \leq q} \hat{A}_{N}^{M}(p,j(q))[\boldsymbol{\phi},\boldsymbol{\theta}]\biggr) \hat{B}_{N}^{M}[\boldsymbol{\Phi}],
\ee
where
\be
\hat{A}_{N}^{M}(p,j(q))[\boldsymbol{\phi},\boldsymbol{\theta}] \equiv  \exp \left(\ui \mathtt{J}_{z}(3)\phi_{(p-1)+j(q)} \right)\exp \left(\ui \mathtt{J}_{y}(1,p) \theta_{(p-1)+j(q)} \right),
\ee
\be
\hat{B}_{N}^{M}[\boldsymbol{\Phi}] \equiv \prod_{1 \leq c \leq N-1} \exp \left(\ui \mathtt{J}_{z}([c+1]^2-1)\Phi_{(N(N-1)/2) + c} \right),
\ee
and $j(q) =0$ for $q = N$ with $j(q)=\sum_{i=1}^{N-q}(N-i)$ for $q \neq N$.  
For example, for $N = 4$ and $M = 1$~\Eq{eq:suN} yields (via Appendix~\ref{AppA}) the operator $\hat{U}_{4}^{1}(\boldsymbol{\phi},\boldsymbol{\theta}, \boldsymbol{\Phi})$ that parametrizes the group \SU{4} in the fundamental representation~\cite{Tilma1}:
\begin{align}
\hat{U}_{4}^{1}(\boldsymbol{\phi},\boldsymbol{\theta}, \boldsymbol{\Phi}) =&
\exp \left(\ui \hat{\mathsf{J}}_4^1(3) \phi_1 \right) \exp \left(\ui \hat{\mathsf{J}}_4^1(2) \theta_1 \right) \exp \left(\ui \hat{\mathsf{J}}_4^1(3) \phi_2 \right) \exp \left(\ui \hat{\mathsf{J}}_4^1(5) \theta_2 \right)\exp \left(\ui \hat{\mathsf{J}}_4^1(3) \phi_3 \right) \exp \left(\ui \hat{\mathsf{J}}_4^1(10) \theta_3 \right) \\ \nonumber
&\exp \left(\ui \hat{\mathsf{J}}_4^1(3) \phi_4 \right) \exp \left(\ui \hat{\mathsf{J}}_4^1(2) \theta_4 \right) \exp \left(\ui \hat{\mathsf{J}}_4^1(3) \phi_5 \right) \exp \left(\ui \hat{\mathsf{J}}_4^1(5) \theta_5 \right) \exp \left(\ui \hat{\mathsf{J}}_4^1(3) \phi_6 \right) \exp \left(\ui \hat{\mathsf{J}}_3^1(2) \theta_6 \right) \\ \nonumber
&\exp \left(\ui \hat{\mathsf{J}}_4^1(3) \Phi_1 \right) \exp \left(\ui \hat{\mathsf{J}}_4^1(8) \Phi_2 \right) \exp \left(\ui \hat{\mathsf{J}}_4^1(15) \Phi_3 \right).
\end{align}
Furthermore, for $N = 2$ and $M = 3$ we get 
\be
\hat{U}_{2}^{3}(\boldsymbol{\phi},\boldsymbol{\theta}, \boldsymbol{\Phi}) = \exp \left(\ui \hat{\mathsf{J}}_2^3(3) \phi_1\right) \exp \left(\ui \hat{\mathsf{J}}_2^3(2) \theta_1\right) \exp \left(\ui \hat{\mathsf{J}}_2^3(3) \Phi_1\right).
\ee
Here, $\hat{\mathsf{J}}_2^3(3)$ is just the $4 \times 4$ version of $\hat{J_z}$ and $\hat{\mathsf{J}}_2^3(2)$ is just the $4 \times 4$ version of $\hat{J_y}$.
In other words, the spin-$3/2$ version of the \SU{2} rotations. 
%given in~\Eq{Eulers}.

%-----------------------------------------------------------------------------
%-----------------------------------------------------------------------------
\section{Normalization Requirements}\label{AppC}

For the Weyl function we have given to be informationally complete, it must reproduce the original Hilbert space operator under integration over the appropriate manifold parametrized by~\Eq{eq:suN}.
Here we will give the volume normalized differential element necessary to integrate any representation of a \SU{N} Wigner or Weyl function, such that
\ba 
\ud\GenWigPar &\rightarrow & \frac{d^M_N}{V_{\CPSpace}} \,\ud V_{\CPSpace}  \label{GenWigVolNorm}\\
\ud\GenWeyPar &\rightarrow & \frac{d^M_N}{V_{\SU{N}}}\,\ud V_{\SU{N}} \label{GenWeyVolNorm}
\ea
which when evaluated for $\mathcal{C}\mathrm{P}^1$ and \SU{2} correspond to~\Eq{WigVolNormD} and~\Eq{WeyVolNormD} respectively when using $d^M_N$ as defined in~\Eq{TheOmega}. 
The difference in volume normalization in~\Eq{GenWigVolNorm} and~\Eq{GenWeyVolNorm} is due to the fact that the Wigner function is defined over the complex projective space in $N-1$ dimensions $\CPSpace$, whereas the Weyl function is defined over the full manifold of $\SU{N}$.

To calculate the invariant volume element for $\CPSpace$ we use the following from~\cite{Tilma2,UandCPN}: 
\ba
\label{eq:dvSUNCP}
\ud V_{\CPSpace} &=& \biggl(\prod_{2 \leq k \leq N} \mathbb{K}(k) \biggr) \ud \phi_{N-1} \ud \theta_{N-1} \ldots \ud \phi_{1} \ud \theta_{1}, \nonumber \\
\mathbb{K}(k) &=& 
\begin{cases}
\sin(2\theta_{1}) \, &k=2, \\
\cos(\theta_{k-1})^{2k-3}\sin(\theta_{k-1}) \, & 2 <k< N, \\
\cos(\theta_{N-1})\sin(\theta_{N-1})^{2 N-3} \,
&k=N,
\end{cases}
\ea 
where the integration is over the following ranges~\cite{Tilma2,UandCPN},
\bel{eq:cpnranges1}
0 \le \phi_{j} \le 2\pi \; \text{and} \; 0 \le \theta_{j} \le \frac{\pi}{2} \; \text{for} \; 1 \le j \le N-1
\ee
such that
\bel{CPNVolume}
V_{\CPSpace} = \int_{\GenWigPar} \ud V_{\CPSpace}. 
\ee

Now considering the \SU{N} volume element, we use the overall volume of the manifold, which does not depend on the dimension of the representation $M$~\cite{UandCPN}.
As such, the volume 
\bel{SUNVolume}
V_{\SU{N}} = \int_{\GenWeyPar} \ud V_{\SU{N}}
\ee
is generated by integrating the invariant integral measure of \SU{N} derived from~\Eq{eq:suN}:
\ba
\ud V_{\SU{N}}&=& \biggl(\prod_{N \geq q \geq 2}\;\prod_{2 \leq p \leq  q}\mathtt{Ker}(p,j(q))\biggr)\ud \boldsymbol{\phi} \ud \boldsymbol{\theta} \ud \boldsymbol{\Phi}, \nonumber  \\
\mathtt{Ker}(p,j(q))&=&
\begin{cases}
\sin(2\theta_{1+j(q)}) \, &p=2, \\
\cos(\theta_{(p-1)+j(q)})^{2p-3}\sin(\theta_{(p-1)+j(q)}) \, &2<p<q, \\
\cos(\theta_{(q-1)+j(q)})\sin(\theta_{(q-1)+j(q)})^{2q-3} \, &p = q,
\end{cases}
\label{eq:dvSUN}
\ea
and $j(q)$ is from~\Eq{eq:suN}. 
The method for the generating the ranges of integration for the full volume of \SU{N} are given in~\cite{Tilma2}. 
For completeness, we note that it has been shown~\cite{MByrdp1,MByrdp2,Tilma1,Tilma2} that the above is mathematically equivalent to the Haar measure~\cite{Marinov,Marinov2} for \SU{N}.

It is important to note that the integration ranges for the calculation of~\Eq{SUNVolume} are equivalent to those used to calculate~\Eq{CPNVolume} but are not equal.
While the ranges of integration for the ``local rotations'' $\theta_{j}$ do not change, the ranges of integration for the ``local phases'' $\phi_{j}$ and the ``global phases'' $\Phi_{j}$ used in the calculation of the overall volume $V_{\SU{N}}$ are modified from those used to calculate $V_{\mathcal{C}\text{P}^{N-1}}$.
For example, the ranges needed to calculate $V_{\SU{4}}$ are (from~\cite{Tilma2})
\begin{gather}
0 \le \phi_1,\phi_4,\phi_6 \le \pi \; \text{and} \; 0 \le \phi_2,\phi_3,\phi_5 \le 2\pi, \nonumber \\
0 \le \theta_1, \theta_2, \theta_3, \theta_4, \theta_5, \theta_6 \le \frac{\pi}{2},\nonumber \\
0 \le \Phi_{1} \le 2\pi, \quad 0 \le \Phi_{2} \le 3\pi \left(\frac{1}{\sqrt{3}}\right), \quad 0 \le \Phi_{3} \le 4\pi \left(\frac{1}{\sqrt{6}}\right).
\end{gather}
These ranges yield both a covering of $\SU{4}$~\cite{Tilma1,UandCPN}, as well as the correct group volume for $\SU{4}$~\cite{Marinov,Marinov2}.
One can use these modified ranges to calculate the equivalent version of~\Eq{CPNVolume} for $N=4$ but then the normalization in front of~\Eq{GenWigVolNorm} would have to be changed.

%-----------------------------------------------------------------------------
\end{widetext}
%-----------------------------------------------------------------------------
%-----------------------------------------------------------------------------
%
% ***** Bibliography *****
%
\bibliographystyle{apsrev4-1}
\bibliography{refs}  

\end{document}